\documentstyle[preprint,prb,aps]{revtex}
\begin{document}
\draft

\begin{title}
{Symmetry-breaking skyrmion states in fractional quantum Hall systems }
\end{title}

\author{ Kang-Hun Ahn and K. J. Chang }
\address{     
Department of Physics, Korea Advanced Institute of Science and Technology,\\
  373-1 Kusung-dong, Yusung-ku, Taejon, Korea 
          }
\maketitle

\date{\today}
\widetext
\begin{abstract}
We calculate in an analyical fashion the energies and net spins of
skyrmions in fractional quantum Hall systems, based on the suggestion that
skyrmion states are spontaneously $L_Z$ and $S_Z$ symmetry-breaking states.
The quasihole-skyrmion state with a charge $-e/3$ around $\nu$ = 1/3,
where the ground state is known as a spin-polarized ferromagnetic state,
is found to exist even in high magnetic fields up to about 7 T for
GaAs samples.
\end{abstract}
\pacs{ }

\narrowtext
\newpage

Since recent observations of a charged spin-texture in the quantum Hall 
system, i.e., skyrmion, an instanton with a topological twist in spin
space has been the subject of great interest. \cite{barret,schmeller,aifer}
The effective field theory of nonlinear sigma (NL$\sigma$) model
predicted that relevant excitations near $\nu = 1/M$ with odd integer $M$
are skyrmions with a macroscopic spin reversal in the limit of weak Zeeman
coupling.\cite{sondhi}
However, the NL$\sigma$ model has not been successful in describing the
observed net spin of skyrmion in the presence of Zeeman coupling.
\cite{barret,schmeller,aifer,sondhi}
There have been several microscopic theoretical attempts to overcome
the drawback of the field theory. \cite{fertig,nayak,mac2,kamilla}
In those microscopic theories, since the Hamiltonian is invariant
under rotations in both the spin and coordinate spaces, skyrmion states
have been considered as eigenstates of $L_Z$ and $S_Z$,
where $L_Z$ and $S_Z$ denote the z-components of the total orbital
and spin angular momentums, respectively.
Recent finite-size calculations around $\nu$ = 1 by Xie and He, \cite{xie2}
however, showed almost degenerate states with different spins that may
give rise to the symmetry-breaking skyrmion state.
This calculational results lead us to consider that the skyrmion states
are spontaneously symmetry breaking states for weak Zeeman couplings.

In this work, we describe skyrmions in a microscopic manner, which are 
defined as spontaneously symmetry-breaking ground states, and calculate in an
analytical fashion the total energies and net spins of skyrmion 
excitations at fractional filling factors $\nu$ =
1, 4/5, 2/3, 1/3, and 1/5 in infinite systems.
The theoretical approach \cite{ahn1} used here is a generalization of
the Hartree-Fock calculation by Fertig and his coworkers \cite{fertig}
around $\nu$ = 1, and it can be used for systems with any rational filling
factors through finite-size calculations.
We find that the skyrmion near $\nu$ = 1/3 is a relevant quasiparticle
state even in very high magnetic fields, contrary to the results of
Kamilla, Wu, and Jain, \cite{kamilla} who treated the skyrmion at $\nu$=1/3 
in terms of composite fermions. \cite{jain}
Near $\nu$ = 1, we find that the symmetry-breaking state is good approximate
to the quantized skyrmion state for high magnetic fields, close to a critical
value where a transition into a spin-polarized quasihole state occurs,
although its energy is higher than for the qunatized skyrmion states with
1 and 2 reversed spins in this field region.
As the electron correlation and Zeeman coupling increase,
the asymmetry between the quasihole-skyrmion and quasiparticle-skyrmion
states is found to be significant, while this feature was not explained
in the effective field theory.

We begin with a conjecture that the skyrmion state is invariant under
a particular rotation $R_{\theta} \equiv \exp [i\theta (L_Z - S_Z )]$. 
This conjecture is consistent with the result of the $\sigma$ model
\cite{rajarman} and was verified by recent numerical calculations 
around $\nu$ = 1. \cite{xie2}
The rotational invariance under $R_{\theta}$ is not associated with
the exact symmetry of the Hamiltonian, but originated from the spin-charge
relation in the lowest Landau level. \cite{moon}
Then, skyrmions must be eigenstates of $L_Z-S_Z$, although it is not
necessary to be eigenstates of $L_Z$ and $S_Z$.
Considering the local spin configuration in the NL$\sigma$ model,
\cite{sondhi,rajarman} the other type of skyrmions are similarily
required to be eigenstates of $L_Z+S_Z$.
As it will be shown that the skyrmion with $L_Z - S_Z$ symmetry 
has a charge  $- \nu e$, we call it quasihole(QH)-skyrmion, while
the skyrmion with $L_Z + S_Z$ symmetry and a charge $+ \nu e$ is called
quasiparticle(QP)-skyrmion, following the notations of
Ref. \protect\onlinecite{kamilla}.

To construct the variational many-body wavefunction for a skyrmion
near fractional filling factor $\nu$ in a plane geometry, we consider
the incompressible spin-polarized ground state $\Phi^{\nu}$ of $N$ electrons
at this filling factor in the symmetric guage,
\begin{eqnarray}
\Phi^{\nu} = \sum_{i_{1} < i_{2} < ... < i_{N} }
C^{\nu} (i_{1} , i_{2} , ..., i_{N} ) \prod_{k=1}^{N}
c^{ \dagger }_{i_k, \uparrow} \mid 0 \rangle,
\end{eqnarray}
where $c_{m, \sigma}^{\dagger}$ creates an electron of spin $\sigma$ and
angular momentum $m$ in the lowest Landau level, and the coefficients
$C^{\nu}(i_1,i_2,...,i_N)$ can be obtained by the exact diagonalization
of the Hamiltonian or by expanding the Laughlin functions. 
Then, the microscopic wave function for a single 
QH-skyrmion ($\Psi_{-}^{\nu}$)
or an 
QP-skyrmion ($\Psi_{+}^{\nu}$) is expressed similarly,
\begin{eqnarray}
\Psi_{\pm}^{\nu} = \sum_{i_{1} < i_{2} < ... < i_{N} }
D^{\nu,\pm} (i_{1} , i_{2} , ..., i_{N} ) \prod_{k=1}^{N}
\gamma^{ \dagger }_{i_k,\pm} \mid 0 \rangle,
\label{skfunc}
\end{eqnarray}
where $\gamma^{\dagger}_{m,-}$ ($\gamma^{\dagger}_{m,+}$) creates a
single electron in the QH-skyrmion (QP-skyrmion) state.
For the QH-skyrmion and QP-skyrmion states to be eigenstates of $L_Z-S_Z$
and $L_Z+S_Z$, respectively, the creation operators must satisfy the
relation $\gamma^{\dagger}_{m,\pm} = u_{m} c_{m, \downarrow}^{\dagger}
+ v_{m} c_{m\mp 1, \uparrow}^{\dagger}$, where $u_m$ and $v_m$ are variational
parameters with the condition $|u_{m}|^{2} +|v_{m}|^{2}$ = 1.
Assuming that QH-skyrmion and QP-skyrmion excitations are localized near
the origin, their wave functions $\Psi_{\pm}^{\nu}$ will be similar to
$\Phi^{\nu}$ at sufficiently large distances from the origin.
In a spin-polarized configuration with $v_m$ = 0 and $u_m$ = 1 for
$m$ = 0, 1, 2,..., the coefficients $D^{\nu,\pm}$ must be equivalent to
$C^{\nu}$ because the Coulomb interactions between up-spin electrons
are the same as those for down-spin electrons.
Thus, we decide to approximate $D^{\nu,\pm}$ by $C^{\nu}$ for a proper
description of electron correlations between reversed-spin electrons.
It is noted that for $\nu =1 $ the QH-skyrmion and QP-skyrmion wavefunctions
are consistent with those proposed by Fertig and his coworkers. \cite{fertig}
If $u_m$ = 0 for all $m$'s, the QH-skyrmion state becomes the quasihole state
of MacDonald and Girvin. \cite{macdonald}
In the QP-skyrmion state, since the single particle state
with angular momentum $m$ = -1, which belongs to the higher Landau level,
is involved, it must be projected onto the lowest Landau level by setting
$u_0$ = -1 and $v_0$ = 0.

The Hamiltonian at filling factor $\nu$ is written as  
\begin{eqnarray}
\nonumber
&H& = - \frac{1}{2}g^{*} \mu_B B \sum_{m} [
c_{m,\uparrow}^{\dagger}
c_{m,\uparrow} -
c_{m,\downarrow}^{\dagger}
c_{m,\downarrow} ]
\\ \nonumber
&&-\nu \sum_{m,n}V_{m n m n} [
c_{m,\uparrow}^{\dagger}
c_{m,\uparrow}+ 
c_{m,\downarrow}^{\dagger}
c_{m,\downarrow} - \nu /2]
\\
&+&\frac{1}{2} \sum_{m_1m_2m_3m_4}
V_{m_1m_2m_3m_4}^{\pm}
\gamma_{m_1,\pm}^{\dagger}
\gamma_{m_2,\pm}^{\dagger}
\gamma_{m_4,\pm}
\gamma_{m_3,\pm}.
\label{hamilt}
\end{eqnarray} 
Here $V^{\pm}_{m_1 m_2 m_3 m_4}$ represents the electron-electron
interaction,
\begin{eqnarray}
V^{\pm}_{m_1 m_2 m_3 m_4} &=& \int d^{2} q_1 d^{2} q_2
\psi^{\dagger}_{m_1,\pm}(\vec{q_1})
\psi^{\dagger}_{m_2,\pm}(\vec{q_2})
\nonumber \\ & \times &
\frac{e^{2}}{\epsilon |\vec{r_1} -\vec{r_2}|}
\psi_{m_3,\pm}(\vec{q_1})
\psi_{m_4,\pm}(\vec{q_2}),
\label{el-el}
\end{eqnarray}
where $\psi_{m,\pm}$ denote the single-particle states created by
$\gamma_{m,\pm}^{\dagger}$ and the integrations are performed over both spin
and space variables.
In the second term which gives the Coulomb interactions between electrons
and uniform neutralizing background charges and the repulsive interactions
between the neutralizing charges, $V_{m_1 m_2 m_3 m_4}$ is defined similarly
to $V^{\pm}_{m_1 m_2 m_3 m_4}$, by replacing $\psi_{m,\pm}$ with
the states $\phi_{m,\uparrow}$ created by $c_{m,\uparrow}^{\dagger}$. 
The expectation values of the Hamiltonian for the QH-skyrmion and QP-skyrmion
states $\Psi^{\nu}_{\pm}$ can be written analytically as functions of $u_m$
and $v_m$, using the following relations;
\begin{eqnarray}
\nonumber
\langle \Psi_{\pm}^{\nu} &|&
c^{\dagger}_{m,\downarrow} c_{m,\downarrow}
| \Psi_{\pm}^{\nu} \rangle =
|u_m |^2 \langle \Phi^{\nu} | c^{\dagger}_{m,\uparrow} c_{m,\uparrow}
| \Phi^{\nu} \rangle 
\\  
\nonumber
\langle \Psi_{\pm}^{\nu} &|&
c^{\dagger}_{m,\uparrow} c_{m,\uparrow}
| \Psi_{\pm}^{\nu} \rangle =
|v_{m \pm 1} |^2 \langle \Phi^{\nu} | c^{\dagger}_{m\pm 1,\uparrow} 
c_{m\pm 1,\uparrow} | \Phi^{\nu} \rangle 
\\
 \langle \Psi^{\nu}_{\pm} &|& \gamma_{m_1,\pm}^{\dagger}
\gamma_{m_2,\pm}^{\dagger} \gamma_{m_4,\pm} \gamma_{m_3,\pm}
| \Psi^{\nu}_{\pm} \rangle 
\nonumber \\ &=&
 \langle \Phi^{\nu} | c_{m_1,\uparrow}^{\dagger}
c_{m_2,\uparrow}^{\dagger} c_{m_4,\uparrow} c_{m_3,\uparrow}
| \Phi^{\nu} \rangle.
\label{expectation}
\end{eqnarray}
In the case of $\nu=1/M$, the analytic forms of the expectation values of the
products of creation and annihiliation operators on the right hand side of
Eq. (\ref{expectation}) are given in literatures. \cite{macdonald,analytic}
With the particle-hole symmetry relation,\cite{p-hole}
\begin{eqnarray}
\langle \Phi^{1/M}&|&c^{\dagger}_{m_1,\uparrow}c^{\dagger}_{m_2,\uparrow}
c_{m_4,\uparrow}c_{m_3,\uparrow}|
  \Phi^{1/M} \rangle
\nonumber \\   &=&
\langle \Phi^{1-1/M} | c_{m_1, \uparrow}c_{m_2, \uparrow}
c^{\dagger}_{m_4, \uparrow}c^{\dagger}_{m_3, \uparrow} |
  \Phi^{1-1/M} \rangle,
\label{symmetry}
\end{eqnarray}
we can extend the analytic expression to the $\nu = 1 - 1/M$ state.
The net spin and energy of the stable skyrmion for a given Zeeman coupling
are determined by minimizing the total energy with respect to $u_m$ and
$v_{m}$, which is accomplished by the conjugate gradient method. \cite{recipe}

For $\nu$ = 1, 4/5, 2/3, 1/3, and 1/5, the energy
differences ($\delta \epsilon_{-}$) between the QH-skyrmion 
and the spin-polarized
quasihole state defined as $\Psi^{\nu}_{0 -} \equiv \Psi_{-}^{\nu}
(u_m = 0; m = 0, 1, 2,...)$ are plotted in Fig. 1(a), as a function of
the effective Zeeman coupling $\tilde{g}$ [= $\frac{1}{2}
{g^{*} \mu_{B} B}/(e^{2}/\epsilon l)]$,
where $l$ is the magnetic length $\sqrt{ \hbar c/(e B)}$.
We find that QH-skyrmions are relevant excitations around the filling
factors considered here, provided that the ground states are maximally
polarized at these filling factors.
Near $\nu$ = 1/3 and 1/5, the maximum magnetic fields for the existence
of QH-skyrmions are estimated to be 7.2 and 1.3 T, respectively, and these
values are much higher than the results of Kamilla, Wu, and Jain;\cite{kamilla}
their Monte Carlo calulations showed the corresponding values of 1.6 and 0.21 T.
In this case, we choose the effective $g$-factor of $g^{*}$ = 0.44,
the effective mass of $m^{*}$ = 0.067 $m_{e}$, and the dielectric
constant of $\epsilon$ = 13 for GaAs.
The discrepancies between the two calculations are mainly caused
by the fact that since the energies of our symmetry breaking states are
variationally minmized, these are relatively lower than the results
of Ref. \protect\onlinecite{kamilla}.

Our results indicate that the QH-skyrmions around $\nu$ = 2/3 and 4/5 are 
reliable quasiparticle states, if the ground states at these filling
factors are the particle-hole conjugate states of the Laughlin
$\nu$ = 1/3 and 1/5 states.
In fact, the observed quantum Hall effect at $\nu$ = 2/3 implies the
possibility of the skyrmion excitation.
Here we point out that both the QH- and QP-skyrmion states around $\nu$ =
2/3 can not be explained in the composite fermion picture, \cite{jain}
where the $\nu$ = 2/3 state is obtained from the $\nu$ = 2 singlet state
by composite fermion transformation.
Since the ground state at $\nu$ = 2/3 is a spin-singlet state in the
absence of Zeeman coupling, \cite{xie1} a lower bound for $\tilde{g}$
would exist in the skyrmion excitation.
Using the calculated energies of the polarized and unpolarized ground
states at $\nu$ = 2/3, \cite{xie1} we find that $\tilde{g}$ should be
greater than 0.0087.
There are some experimental evidences for the skyrmion excitation at
$\nu$ = 2/3; the magnetoabsorption spectroscopy reveals a dramatic
reduction of the spin polarization near $\nu$ = 2/3, \cite{eisen} and the
magnetotransport measurement of the activation energy indicates that relevant
quasiparticles are associated with several reversed spins. \cite{aifer}

The QP-skyrmion excitations are also examined and found to be
energetically more stable, as compared to their reference states defined as
$\Psi^{\nu}_{0 +} \equiv \Psi_{+}^{\nu}(u_0= -1, u_m = 0; m = 1, 2,...)$,
as shown in Fig. 1(b).
However, these excitations are less stable than the spin-polarized states
$\Psi_{+}^{\nu}(u_m = 0; m = 1, 2,...)$ with $\gamma^{\dagger}_{0,+}$
replaced by $c_{0, \uparrow}^{\dagger}$ for stronger Zeeman couplings,
for examples, for $\tilde{g} >$ 0.032 and 0.017 at $\nu$ = 1/3 and 1/5,
respectively.
Thus, the QP-skyrmions can not be relevant quasiparticles in the regime
of strong Zeeman coupling, and the neutral excitation might be 
a pair of QH-skyrmion and quasielectron.

Fig. 1(c) shows the calculated spins $\delta s_{-}$ and $\delta s_{+}$
of the QH-skyrmion and QP-skyrmion states relative to their reference
states, respectively.
In the limit $\tilde{g} \rightarrow 0$, both $\delta s_{-}$ and $\delta s_{+}$
are found to increase and become similar, consistent with the effective
field theory.
However, for stronger Zeeman couplings which increase time-reversal
symmetry breaking, the difference between $\delta s_{-}$ and $\delta s_{+}$
becomes more significant, and similar behavior is also found for lower
filling factors, where the electron correlations are stronger.
The prominent difference between the QH-skyrmion and QP-skyrmion states
at fractional filling factors ($\nu <1$ ) is the existence of the critical value
for $\tilde{g}$, at which the QH-skyrmion undergoes a transition into
the spin-polarized quasihole state $\Psi_{0 -}^{\nu}$.
As the Zeeman coupling becomes stronger, since the energy of the reversed
electrons also increases, the QH-skyrmion size tends to shrink to zero at
$\tilde{g}_c$, resulting in the stable spin-polarized quasihole state.
However, we do not find a similar transition in the QP-skyrmion state for
$\nu < 1$,
as discussed earlier.
For filling factors $\nu$ = 1, 4/5, 2/3, 1/3, and 1/5, the calculated
values for $\tilde{g_c}$ are found to be close to 0.0265, 0.0207, 0.0171,
0.0079, and 0.0034, respectively.
For $\tilde{g} > \tilde{g}_c$, all $u_m$'s for the QH-skyrmion state
are found to be zero.
This critical behavior is well described by the known Landau
theory with the order parameter $u_m$ and the temperature replaced with
$\tilde{g}$.
In the Landau theory, the rotational symmetry of a ferromagnetic system
is spontaneously broken below the critical temperature.
Similarly, our quantum Hall system has the spontaneously broken rotational
symmetry below the critical effective Zeeman coupling $\tilde{g_c}$.
Then, in the limit $ \tilde{g} \rightarrow \tilde{g_c}^{-}$,
the number of reversed spins, i.e., the order parameter, is expressed such as
\begin{eqnarray}
\delta s_{-} = \nu   \sum_{m=0}^{m=\infty} |u_m|^2 
\sim  |\tilde{g}-\tilde{g}_c|^{\beta},
\end{eqnarray}
where $\beta$ is the critical exponent.
From the calculated net spins $\delta s_{-}$, the value of $\beta$
is estimated to be about 1 and the order parameter exhibits clearly
the scaling behavior, as illustrated in Fig. 1(d).

As $\tilde{g}$ approaches $\tilde{g}_{c}$, in fact, the numbers of the
reversed spins become less than one, as shown in Fig. 1(c), and this feature
is attributed to the variational method used here.
Since the symmetry breaking is a macroscopic phenomena in the thermodynamic
limit, our calculations might be valid for weak Zeeman couplings,
where the number of reversed spins is a macroscopic number.
Thus, we examine precisely the quantized skyrmion states \cite{nayak}
with definite quantum numbers for $S_Z$ and $L_Z$ near $\tilde{g}_{c}$,
using the variational energy functions driven here, which allow for
analytical calculations, as Bayman did for the superconducting BCS
wavefunctions. \cite{bayman}
Near $\nu$ = 1, the relative energies $\delta \epsilon_{-}$ of the
quantized QH-skyrmion states with the numbers of reversed spins $s$ = 1
and 2 are calulated to be $2 \tilde{g}-0.056$ and $4 \tilde{g}-0.082$,
respectively, and found to be lower than that of the symmetry-breaking state.
Since the resulting critical values for $\tilde{g}$ are similar to that
of the symmetry-breaking state, our symmetry-breaking wavefunctions are
considered to be good approximate to the quantized skyrmion states near
$\tilde{g}_{c}$.

The charge and spin densities of QH-skyrmion and QP-skyrmion, which are
defined by the expectation values of
$\hat{\rho}_{s}({\bf r}) = \frac{1}{2}\sum_{m} |\phi_m ({\bf r})|^{2}
 (c_{m,\uparrow}^{\dagger} c_{m,\uparrow}
 -c_{m,\downarrow}^{\dagger} c_{m,\downarrow})$ and
$\hat{\rho}({\bf r}) = \sum_{m} |\phi_m ({\bf r})|^{2}
 (c_{m,\uparrow}^{\dagger} c_{m,\uparrow}
 +c_{m,\downarrow}^{\dagger} c_{m,\downarrow})$, respectively, are drawn
for a Zeeman coupling constant of $\tilde{g}$ = 0.005
at $\nu$ = 1/3 and 2/3 in Fig. 2.
Our caculated spin density profile shows that the reversed spins are
truly bound to the skyrmion core, in contrast to previous calculations
for the hard-core potentials.\cite{mac2,kamilla}
It is interesting to note that although the QH-skyrmion and QP-skyrmion
states are constructed in a different way, they have almost the same
spin density profiles, as shown in Fig. 2(a).
This feature may be understood in terms of the effective field theory
that gives exactly the same z-component of the local spin
for the two excitations. \cite{sondhi}
The charge density of the QH-skyrmion exhibits a deficiency in the core region,
while a similar distribution of excess charges is found for the
QP-skyrmion [see Fig. 2(b)], as expected from the effective field theory. 
For all the filling factors considered here, we find that
the changes of the charge densities $\rho ({\bf r})$ associated with
$\Psi_{\pm}^{\nu}$ in the core region is exactly related to the topological
charges $Q = \pm 1$ \cite{moon} via $ \int d^2 {\bf r} ( \rho ({\bf r})
- \nu / (2\pi l^2) ) = \nu Q$,
where $Q$ = 1 for the QP-skyrmion and $Q$ = -1 for the QH-skyrmion excitation,
in good agreement with the well-known conjecture. \cite{moon}

In summary, we have made a microscopic description of skyrmions, which
requires that the skyrmion states are both $L_Z$ and $S_Z$
symmetry-breaking states.
It is suggested that these states are originated from the degenerate
states with different spins, however, they have the same eigenvalues of
$L_Z-S_Z$ for the QH-skyrmion, while $L_Z + S_Z$ for the QP-skyrmion states.
Testing the $\nu \approx$ 1/3, 1/5, 2/3, and 4/5 states,
we find that the difference of the net spins for the QH-skyrmion and
QP-skyrmion excitations becomes significant as the electron correlation
and Zeeman coupling are stronger.
The possibility of skyrmion excitations around $\nu$ = 2/3 is predicted,
which is not described by the composite-fermion model.

This work was supported by the CTPC and CMS at KAIST.
One of the authors (Ahn) would like to thank J. P. Dziarmaga, S.-R. Eric Yang,
J.-J. Kim, Y.-G. Jin, and H.-S. Sim for helpful discussions.

\begin{figure}
\caption{ (a)-(b) Relative energies of the QH-skyrmion ($\delta \epsilon_{-}$)
and QP-skyrmion ($\delta \epsilon_{+}$) states vs effective Zeeman
coupling. (c) Net spins for the QH-skyrmions ($\delta s_{-}$)
and QP-skyrmions ($\delta s_{+}$) states are compared. 
Note that $\delta s_{+}$ is larger than $\delta s_{-}$ for
$\nu < 1$, while they are equivalent at $\nu$ = 1.
(d) Scaled $\delta s_{-}$ vs scaled $(\tilde{g}_c -\tilde{g})$.
Circles denote calculations and slopes give the critical exponent $\beta$.} 

\begin{figure}
\caption{
(a) Spin and (b) charge density profiles for $\nu$ = 1/3 and 2/3.
Solid and dotted lines represent the QH-skyrmion and QP-skyrmion states,
respectively. }
\label{fig2}
\end{figure}
\label{fig1}
\end{figure}

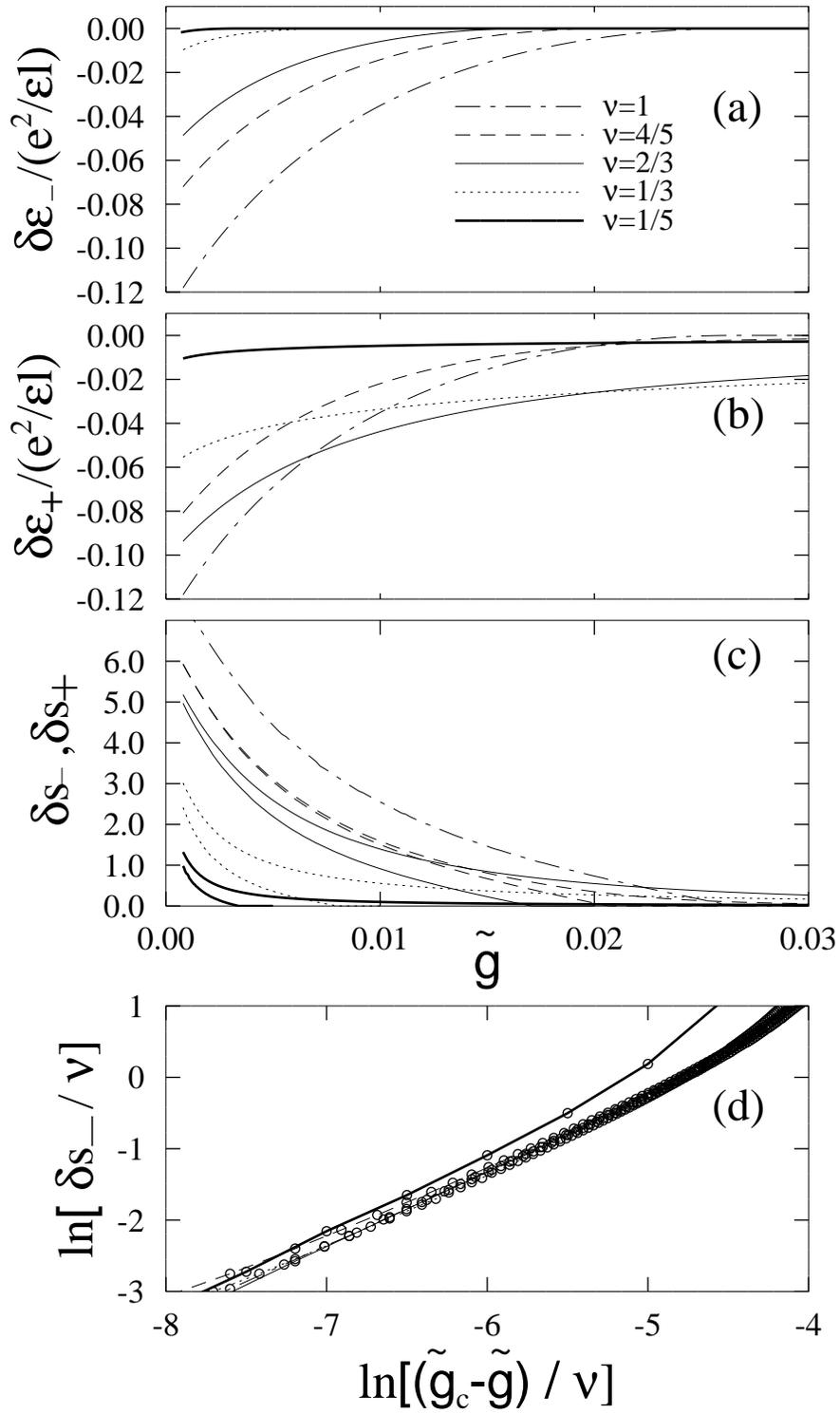
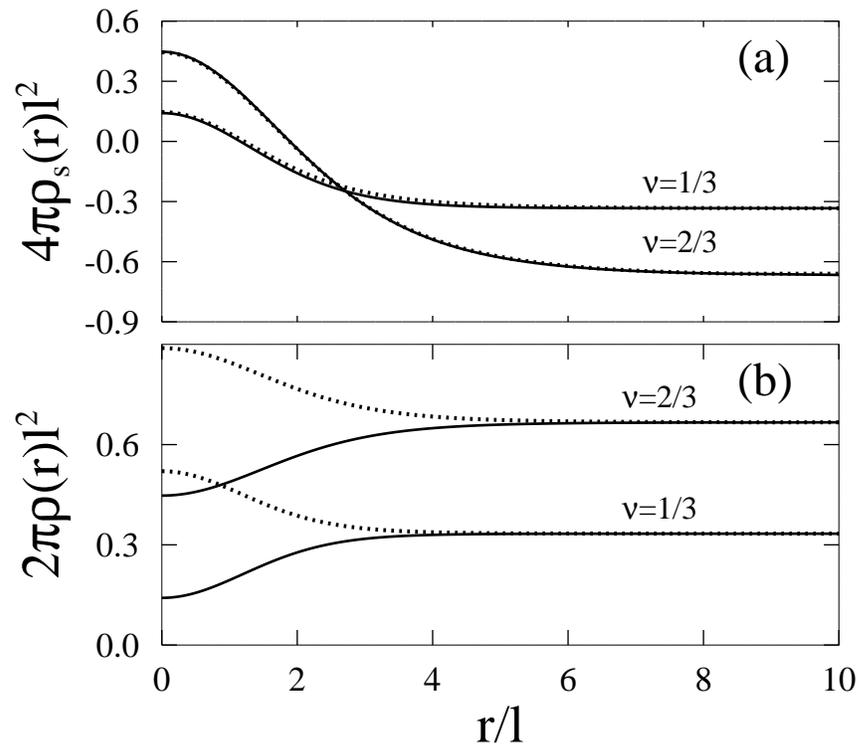
\end{document}